\documentclass[journal=ancac3, manuscript=review]{achemso}

\usepackage{chemformula} % Formula subscripts using \ch{}
\usepackage[T1]{fontenc} % Use modern font encodings

\author{Fan Yang}
\affiliation[ Imperial College London]
{The Blackett Laboratory, Department of Physics, Imperial College London, London SW7 2AZ, United Kingdom}
\email{f.yang16@imperial.ac.uk}

\author{Yao-Ting Wang}
\affiliation[Imperial College London]
{The Blackett Laboratory, Department of Physics, Imperial College London, London SW7 2AZ, United Kingdom}

\author{Paloma Arroyo Huidobro}
\affiliation[ Imperial College London]
{The Blackett Laboratory, Department of Physics, Imperial College London, London SW7 2AZ, United Kingdom}

\author{John B Pendry}
\affiliation[ Imperial College London]
{The Blackett Laboratory, Department of Physics, Imperial College London, London SW7 2AZ, United Kingdom}

\title[An \textsf{achemso} demo]
  {Nonlocal effects in singular plasmonic metasurfaces}

\abbreviations{}

\begin{document}
%\begin{tocentry}
%Some journals require a graphical entry for the Table of Contents.
%This should be laid out ``print ready'' so that the sizing of the
%Inside the \texttt{tocentry} environment, the font used is Helvetica
%8\,pt, as required by \emph{Journal of the American Chemical
%The surrounding frame is 9\,cm by 3.5\,cm, which is the maximum
%permitted for  \emph{Journal of the American Chemical Society}
%graphical table of content entries. The box will not resize if the
%content is too big: instead it will overflow the edge of the box.
%This box and the associated title will always be printed on a
%separate page at the end of the document.
%\end{tocentry}
\begin{abstract}
A local model of the dielectric response of a metal predicts that singular surfaces, such as sharp-edged structures, have a continuous absorption spectrum and extreme concentration of energy at the singularity. Here we show that nonlocality drastically alters this picture: the spectrum is now discrete and energy concentration, though still substantial, is greatly reduced.

\end{abstract}
\section{Introduction}
In the past decade, rapid development in nanofabrication has prompted great interest in plasmonic systems which collect and concentrate light into subwavelength volumes \citep{maier2007plasmonics}. Among these systems, singular plasmonic structures, such as narrow gaps and sharp edges \citep{prodan2003hybridization, hao2004electromagnetic, nordlander2004plasmon, lu2005nanophotonic, romero2006plasmons, aubry2010plasmonic, luo2010surface, pacheco2016description, benz2016single, galiffi2018broadband, urbieta2018atomic, pacheco2018understanding}, can even concentrate the field down to the Coulomb screening length where electron-electron interaction should be considered \citep{garcia2008nonlocal, mortensen2014generalized, toscano2015resonance, fitzgerald2016quantum, ciraci2016quantum}. The hydrodynamical model \citep{boardman1982electromagnetic, ciraci2012probing, moreau2013impact} takes this interaction into account and successfully explains nonlocal effects in the gap of a dimer \citep{fernandez2012PRL, fernandez2012PRB, luo2013surface}. 

In our recent work\citep{pendry2017compacted, yang2018transformation}, we proposed and studied singular metasurfaces within a local description where the optical response of metals is described by a $\omega$-dependent permittivity such as the Drude model, $\varepsilon (\omega) = \varepsilon_{\infty} - \frac{\omega_p^2}{\omega(\omega+i \Gamma)}$. However, the local model assumes that the electron gas is a continuum taking no account of the Fermi surface and the finite density of electrons. In our singular metasurface, the SPPs will propagate to the singularities, where electrons accumulate and reach infinite density, which leads to divergences of the electric field. However, the infinite density is not physical because of the discrete nature of the electron gas, which determines the screening length ($\delta_C \sim 0.1$ nm, for noble metals) \cite{kittel2004introduction} and prevents the electron density from blowing up. In this work, we take this nonlocal effect into account for singular metasurfaces where the scale of the singularity could go below the screening length. Our results show that the singular metasurface is very sensitive to the nonlocal effects of the electron gas and that nonlocality in the metal has to be considered for an accurate description of our singular system. We discuss how singular metasurfaces could be used to reveal nonlocal effects.

\section{Results and discussion}

\begin{figure}[h]
\includegraphics[width=0.6\columnwidth]{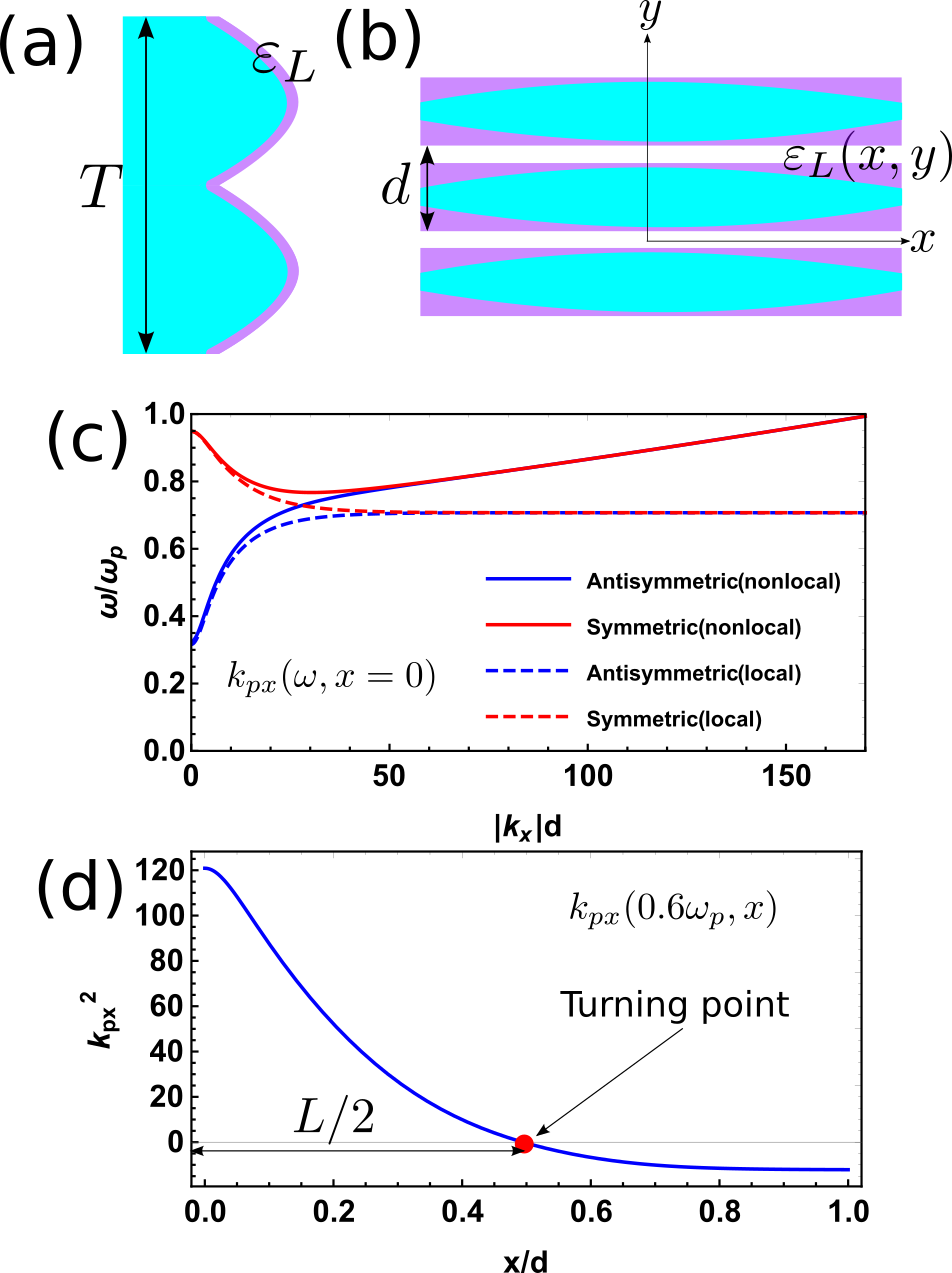}
\centering
\caption{Schematic of nonlocality in the singular metasurface. (a) The screening length of a noble metal (Gold in this paper) smooths the singularity; (b) The decay length of the longitudinal mode is coordinate dependent, due to the $k$-dependence of the longitudinal permittivity $\varepsilon_L$; (c) Dispersion relation at $x=0$ in the slab frame; (d) The coordinate-dependence of $k_{px}^2$ on $x$ for the anti-symmetric mode at $\omega = 0.6 \omega_p$, where the red point marks the turning point for the SPP wave.}
\label{Schematic}
\end{figure}

Fig. \ref{Schematic}(a) is a schematic of our singular metasurface: a periodic array of grooves on a metal surface with sharp edges illuminated by a plane wave. In the presence of nonlocality, we have both transverse and longitudinal modes in the metal. For the transverse mode, the dielectric function is modeled with classical Drude permittivity with $\varepsilon_{\infty}=1$, $\omega_p = 8.95$ eV/$\hbar$ and $\Gamma = 65.8$ meV/$\hbar$ \cite{novotny2012principles}, while for the longitudinal mode, the dielectric function is described within the hydrodynamic model as $\varepsilon_L (\omega, \mathbf{k}) = \varepsilon_{\infty} - \frac{\omega_p^2}{\omega(\omega+i \Gamma) - \beta^2|\mathbf{k}|^2}$ \cite{boardman1982electromagnetic}, where $\beta$ is the nonlocal parameter, measuring the degree of nonlocality. The decay length of the longitudinal mode is $\delta = \frac{\beta}{\sqrt{\omega_p^2-\omega^2}}\approx \frac{\beta}{\omega_p}$ when the frequency is well below the plasmon frequency. In the presence of nonlocality, the surface charge becomes a volume charge with a decay length $\delta$ inside the metal surface, which is depicted as an electron density layer (purple layer) on top of the metasurface with period $T$ shown in Fig. \ref{Schematic}(a). The existence of this layer smooths the singularity and has remarkable effects on the far field spectrum. 

The complex boundary of a singular metasurface complicates the calculation, so we shall map this singular surface into a slab array with period $d$ (see Fig. \ref{Schematic}(b)), where the thickness of the slabs is $d_3$ and the thickness of the dielectric region is $d_1 + d_2$ \citep{yang2018transformation}. In the slab frame, the transverse electric permittivity remains the same since it is $k$-independent and conformal mapping conserves the permittivity in the $x-y$ plane. However, the dependence of the longitudinal permittivity on the $k$ vector complicates the problem because $k$ is not invariant under conformal mapping. Therefore, the longitudinal permittivity in the slab frame becomes coordinate dependent, and is written as 
\begin{equation}
\begin{split}
\varepsilon_L^{z}(\omega, \mathbf{k}, z) 
&=  \varepsilon_L^{z^{'}}(\omega, \left|\frac{d z^{'}}{d z} \right|\mathbf{k}) \\
&=  \varepsilon_{\infty} - \frac{\omega_p^2}{\omega(\omega+i \Gamma)-\frac{\beta^2 d^2 |\mathbf{k}|^2}{T^2} \left| \sinh \left( 2\pi \frac{z}{d}\right)\right|^2} \\
&=  \varepsilon_{\infty} - \frac{\omega_p^2}{\omega(\omega+i \Gamma)-\beta_{eff}^2(z)|\mathbf{k}|^2}
\end{split}
\label{nonlocal permittivity}
\end{equation}
where the primed complex coordinate $z^{'} = x^{'} + i y^{'}$ is for the metasurface frame, while the unprimed coordinate $z = x + i y$ is for the slab frame. The effective nonlocal parameter $\beta_{eff}$ in the slab frame, $\beta_{eff}(z) = \frac{\beta d}{T} \left| \sinh \left( 2\pi \frac{z}{d}\right)\right|$, indicates that the decay length is smaller near the origin and increases when $x \rightarrow \pm \infty$ (see Fig. \ref{Schematic}(b)), i.e. the permittivity is more and more nonlocal along the slab. 

We start by calculating the SPP dispersion in the slab frame. Plane waves (incident, reflected and transmitted waves) in the metasurface frame are modeled as monopole sources in the slab frame and a non-retarded approximation is taken \cite{yang2018transformation}. These three sorts of waves excite the SPPs along the slab. To determine the coefficients of this SPP mode, the tangential component of the magnetic and electric fields ($H_z$ and $E_x$ in our coordinate system) should be matched at the interface between metal and air. However, imposing the continuity of $H_z$ and $E_x$ is not sufficient as there exists an additional longitudinal mode. To solve this indeterminacy, an additional boundary condition, the continuity of $E_y$, is imposed at the interface between metal and air, where $E_y$ is the electric field component normal to the interface. This continuity comes from the assumption of no surface charge in the hydrodynamic model. By imposing these boundary conditions, we can calculate the mode in the slab frame. Since $\beta_{eff}$ is a function of the slab frame coordinate, $k_x$ varies along the slab. To make an analytic solution possible for this inhomogeneity, the WKB approximation is introduced. The WKB approximation, which has been successfully used in plasmonic systems\cite{fernandez2012PRL, fernandez2012PRB, wiener2012nonlocal}, applies when the phase of the SPP wave changes more quickly than its amplitude. With the above additional boundary condition and the WKB approximation, the SPP mode and the corresponding dispersion relation are determined. The dispersion relation for anti-symmetric mode is
{\scriptsize
\begin{equation}
\begin{split}
(\varepsilon-1)\sqrt{k_x^2}(e^{\sqrt{k_x^2}(d_1+d_2)}+1)(e^{\sqrt{k_x^2}d_3}+1) \frac{e^{\kappa d_3}-1}{\kappa(e^{\kappa d_3}+1)} + ((\varepsilon-1)(e^{\sqrt{k_x^2}(d_1+d_2)}-e^{\sqrt{k_x^2}d_3}) + (\varepsilon+1)(e^{\sqrt{k_x^2}d}-1)) = 0
\end{split}
\label{anti-symmetric mode dispersion}
\end{equation}}\normalsize
while the dispersion relation for the symmetric mode is given by
{\scriptsize
\begin{equation}
\begin{split}
(\varepsilon-1)\sqrt{k_x^2}(e^{\sqrt{k_x^2}(d_1+d_2)}-1)(e^{\sqrt{k_x^2}d_3}-1) \frac{e^{\kappa d_3}+1}{\kappa(e^{\kappa d_3}-1)} - ((\varepsilon-1)(e^{\sqrt{k_x^2}(d_1+d_2)}-e^{\sqrt{k_x^2}d_3}) - (\varepsilon+1)(e^{\sqrt{k_x^2}d}-1)) = 0
\end{split}
\label{symmetric mode dispersion}
\end{equation}}\normalsize
where $\kappa = \sqrt{k_x^2 + \frac{\omega_p^2}{\frac{\beta^2 d^2}{T^2} \left| \sinh \left( 2\pi \frac{z}{d}\right)\right|^2}\frac{\varepsilon}{\varepsilon - 1}}$ and $k_{px}$ stands for the plasmon pole. The symmetry of the SPP mode is defined by the parity of $E_x(y)$ in the slab frame: The odd(even) function corresponds to the anti-symmetric(symmetric) mode. The dispersion relation at the center of the slab $k_{px} (\omega, x=0)$ is presented in Fig. \ref{Schematic}(c), together with the local result for comparison \citep{yang2018transformation}. For the local case, both anti-symmetric and symmetric modes asymptotically approach $\omega = \omega_{sp}$. In contrast, within the nonlocal description, this SPP mode asymptotically approaches the longitudinal bulk mode $\omega = \sqrt{\beta^2 k^2 + \omega_p^2}$. In the large $k$ limit, $\omega \approx \beta k$, resulting in a linear dispersion relation, see Fig. \ref{Schematic}(c). In the following, we will focus on the anti-symmetric band which can be excited by a normally incident plane wave. For this anti-symmetric mode, the profile of $k_{px}^2$ on the interface of the slab at a representative frequency within the band $\omega=0.6\omega_p$ is shown in Fig. \ref{Schematic}(d), where the point when $k_{px}^2 = 0$ is called a "turning point" \cite{griffiths2016introduction}. At the turning point, the SPP wave is reflected and the two turning points ($x=-L/2$, $x=L/2$) confine SPP waves to form a Fabry-Perot resonant cavity with width $L$. 

\begin{figure}[h]
\includegraphics[width=0.6\columnwidth]{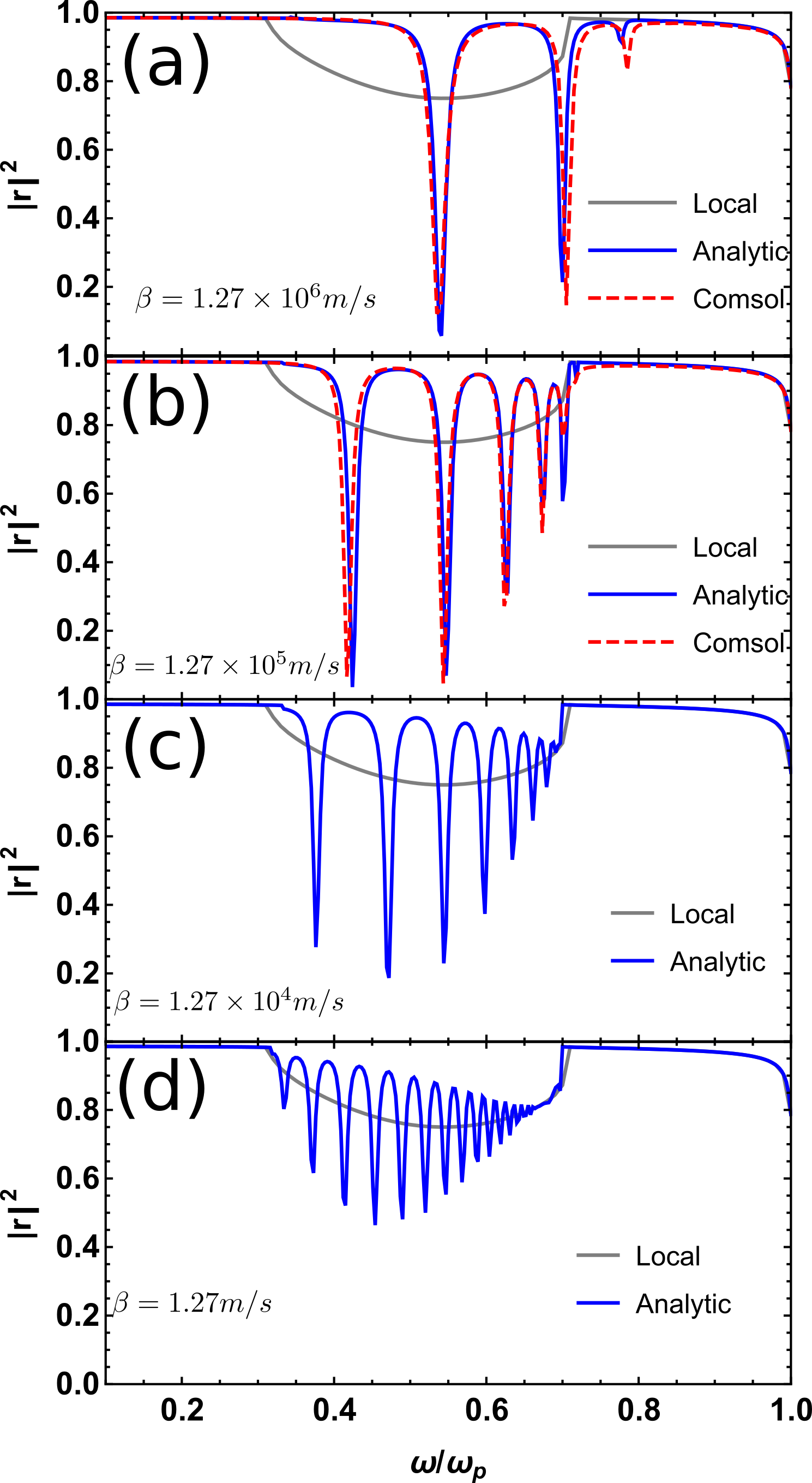}
\centering
\caption{Reflection spectrum of a groove system with $T=10$ nm under normal incidence with different nonlocal parameter $\beta$: (a) $\beta = 1.27\times 10^6$ m/s; (b) $\beta = 1.27 \times 10^5$ m/s; (c) $\beta = 1.27\times 10^4$ m/s; (d) $\beta = 1.27$ m/s. The blue solid line corresponds to analytic calculation while the red dashed line to Comsol simulation, which are compared with a calculation without nonlocality (gray line).}
\label{Reflection_beta}
\end{figure}

With the mode profile in the slab frame (see Methods section for detailed calculations), we can calculate the power flow at $x=0$ and model this power flow as the energy consumption by an effective surface conductivity, which has been previously introduced as flat surface model \citep{yang2018transformation}. This surface conductivity gives us the reflection spectrum. In all the following calculations, the parameter $d_3 = 0.9d$ is used, fixing the shape of the metasurface. 

In Fig. \ref{Reflection_beta} we show how nonlocality greatly affects the optical response of singular metasurfaces. Here we present the reflection spectrum of a groove metasurface with period $T=10$ nm for different nonlocal parameters $\beta$. In Fig. \ref{Reflection_beta}(a), a realistic value of $\beta=1.27\times 10^6$ m/s is used \citep{moreau2013impact}. The theoretical calculation (blue solid line) agrees very well with the simulation results (red dashed line) from the commercial finite element method solver COMSOL MULTIPHYSICS, which are also compared with a local calculation (gray line). It is clear that the spectrum becomes discrete when nonlocality is introduced. To further clarify the effect of nonlocality on the spectrum, we conduct a series of calculations with different $\beta$. When $\beta$ is 10 times smaller ($\beta = 1.27 \times 10^5 $ m/s), more resonances appear in the spectrum, shown in Fig. \ref{Reflection_beta}(b), in which the analytic result also agrees well with numerics. We can reduce $\beta$ further to $1.27\times 10^4$ m/s (Fig. \ref{Reflection_beta}(c)), and even to $1.27$ m/s(Fig. \ref{Reflection_beta}(d)) where Comsol simulations do not converge. The nonlocal spectrum supports more and more peaks as $\beta$ is reduced. In the limit $\beta \rightarrow 0$, the spectrum will become continuous. The origin of the spectrum discretization is that nonlocality blunts the singularity. The resonance condition for different peaks is
\begin{equation}
\begin{split}
\int_{-L/2}^{L/2}k_{px}(x)d x + \phi = 2 n \pi
\end{split}
\end{equation}
where $n$ is an integer and $\phi$ is the phase change at the turning point which is discussed in Methods. The finite range of integration yields a finite number of resonances. As $\beta$ decreases, the turning point is further from the origin and the $k$ integration is larger, thereby allowing for more peaks and appreciably changing the spectrum. In a nutshell, this microscopic feature of nonlocality greatly affects the optical response of the singular metasurface.

\begin{figure}[h]
\includegraphics[width=0.5\columnwidth]{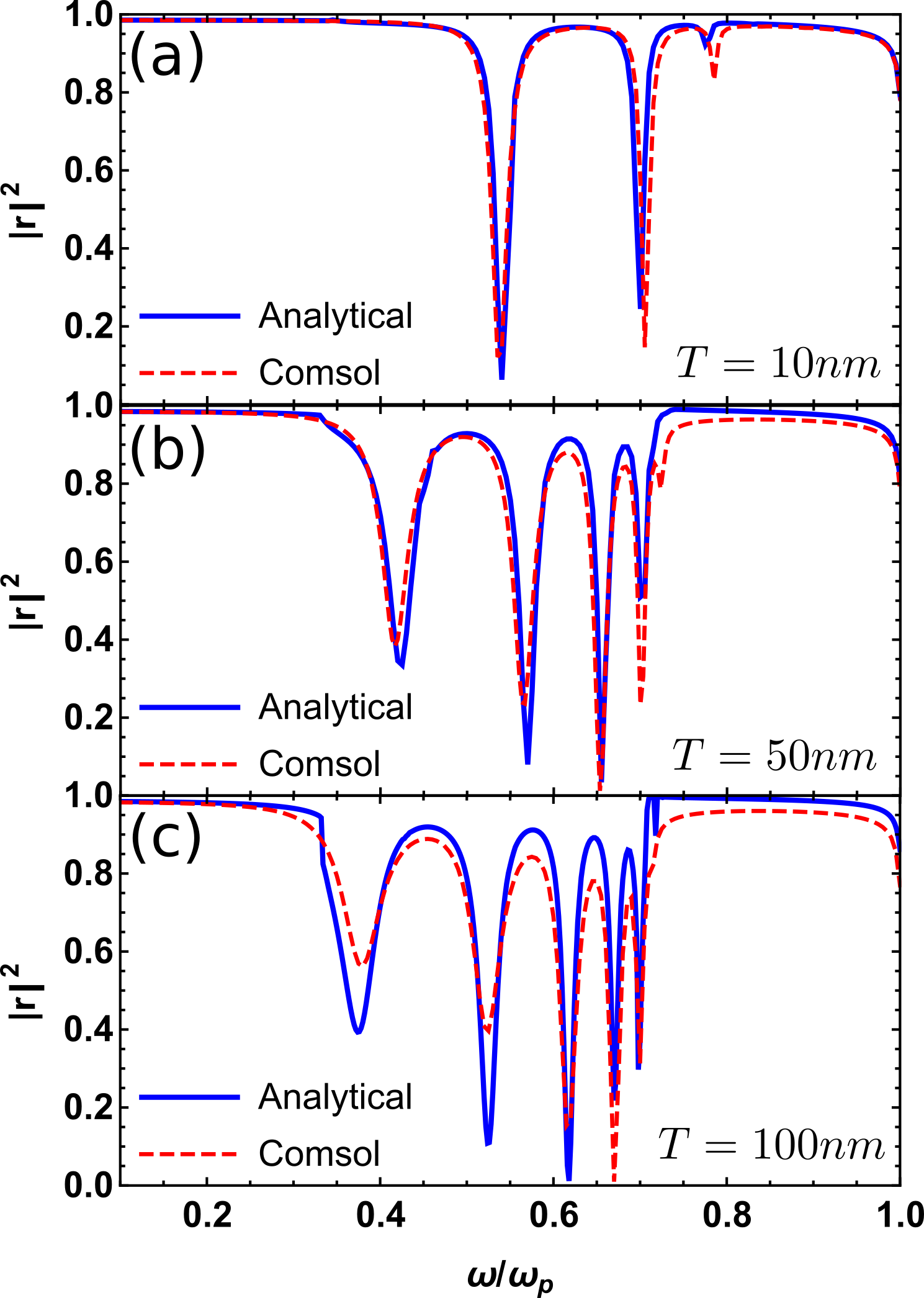}
\centering
\caption{Reflection spectrum of a groove system with $\beta=1.27\times 10^6$ m/s under normal incidence with different period $T$: (a) $T = 10$ nm; (b) $T = 50$ nm; (c) $T = 100$ nm. The blue solid line corresponds to analytic calculation while the red dashed line to Comsol simulation.}
\label{Reflection_T}
\end{figure}

Next, we study the effect of the other length scale of the problem, the metasurface period, while keeping a realistic nonlocal parameter ($\beta=1.27\times 10^6 $ m/s). In Fig. \ref{Reflection_T}(a) we reproduce the reflection spectrum for 10 nm period where we have already seen an excellent agreement with full electrodynamics simulations as in this regime the quasi-static approximation is well satisfied. When the period is increased to 50 nm (Fig. \ref{Reflection_T}(b)), more resonances are excited and at the same time the spectrum moves to lower frequencies. The reason for this is that when the period is increased the length scale where nonlocality is relevant is effectively smaller. Hence, increasing the period has a similar effect to reducing $\beta$ since these are the two length scales which compete in this problem.

Then we further increase the period to 100 nm, where the quasi-static approximation is not expected to be accurate. However, there is still a good agreement between our theory and simulations. In particular, the linewidth is not broadened in comparison with $T= 10$ nm and $T = 50$ nm, except for the low order resonance peaks whose linewidth gets broadened slightly. This is a remarkable difference from localized plasmonic structures of sizes comparable to the period considered here, where radiative broadening is important and a theory beyond quasi-statics is needed to accurately describe these systems\cite{fernandez2012PRL}. Different from localized plasmonic structures, the periodic singular metasurface does not have a strong radiative broadening because by increasing the period we have less periods within overall length. The first resonance peak which has the highest broadening forms a stronger dipole moment, so its broadening is larger than the higher order peaks whose electron distribution is more homogeneous.

\begin{figure}[h]
\includegraphics[width=1\columnwidth]{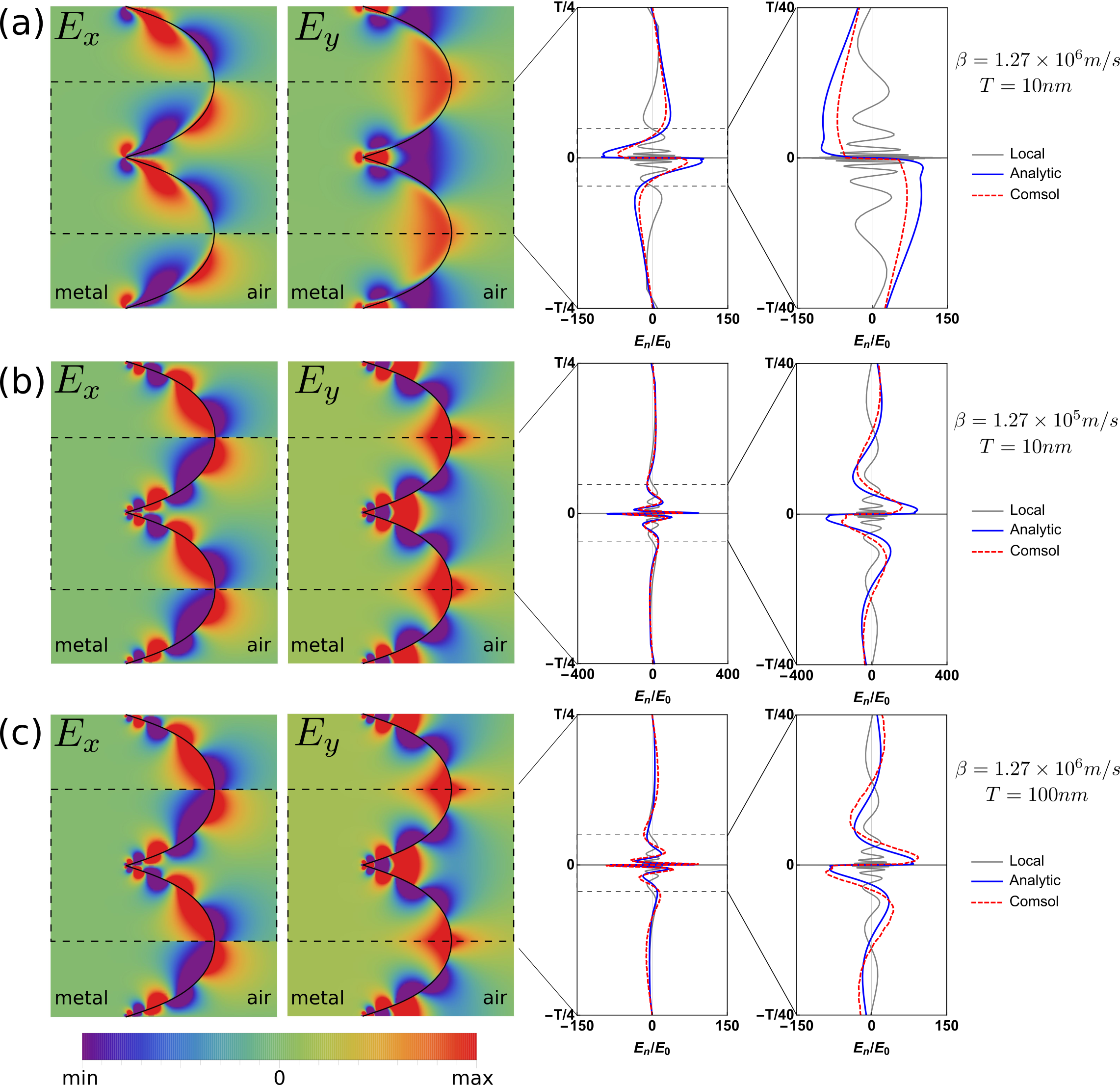}
\centering
\caption{Electric field profile for a groove system under normal incidence: (a) $\beta=1.27\times 10^6$ m/s and $T = 10$ nm; (b) $\beta=1.27\times 10^5$ m/s and $T = 10$ nm; (c) $\beta=1.27\times 10^6$ m/s and $T = 100$ nm. The first two columns are field contour plot for $E_x$ and $E_y$, the last two columns are normal component of electric filed on the interface of metasurface shown with different scales in $y$-axis.}
\label{Efield}
\end{figure}

After studying the far field spectrum, we turn to the near field profile in the presence of nonlocality. Fig. \ref{Efield} illustrates the electric field profile of the metasurface, in which the first and the second column depict the electric field ($E_x$ and $E_y$) distribution in one period ($-T/2 < y <T/2$). Also, the electric field $E_n$ (the component normal to the interface, which dominates for the anti-symmetric mode) along the surface is presented in the last two columns with different scales in the $y$-axis. Fig. \ref{Efield}(a) shows the electric field profile for the second resonance peak in the reflection spectrum of Fig. \ref{Reflection_beta}(a) (period $T=10$ nm, $\beta=1.27 \times 10^6$ m/s), at a frequency of $\omega = 0.7\omega_p$. From this field profile, we see that the electric field is continuous across the interface because the hydrodynamic model assumes no surface charge. On the contrary, the existence of a surface charge in the local description makes the normal component of electric field discontinuous. On the other hand, the field near the singularity is not as localized as the field in the local case. This is because the smearing of the singularity introduced by nonlocality results in a weaker compression of the plasmon wavelength as it travels towards the singularity. This effect causes a strong reduction of the field enhancement in the vicinity of singularity. In fact, while the electric field diverges at the singularity in the local case, it has a value of approximately 70 when nonlocality is taken into account, as shown by the plot of the field at the interface. In addition, the agreement between analytical and numerical results is excellent except near the singularity. This discrepancy between theory and Comsol is due to the WKB approximation, which does not work well when the $k$-vector is small. 

To further unveil the effect of nonlocality on the field enhancement, in Fig. \ref{Efield}(b) we consider a metasurface with the same period as in Fig. \ref{Efield}(a) but with a smaller nonlocal parameter ($\beta=1.27\times 10^5$ m/s). We plot the fields of the fifth resonance peak in Fig. \ref{Reflection_beta}(b), at the same frequency $\omega= 0.7\omega_p$ as Fig. \ref{Efield}(a). Comparing to the case with a realistic description of nonlocality (panel (a)), the field in this case shows more oscillations and presents a sharper change at the metal surface, looking more alike to the local situation. In addition, the plot of the field along the interface demonstrates how the field enhancement strongly increases as compared to panel (a), which is consistent with a divergent field in the limit of $\beta \rightarrow 0$.

Finally, we also consider the near field profiles for a metasurface with a long period ($T=100$ nm) and nonlocal metal (realistic $\beta$). The field profile for the fifth-order resonance in the metasurface (the fifth peak in spectrum in Fig. \ref{Reflection_T}(c), $\omega = 0.7 \omega_p$), is shown in Fig. \ref{Efield}(c). Similar to decreasing the value of $\beta$ for a fixed period, keeping $\beta$ and increasing the period effectively reduces the length scale of nonlocality. Hence, the fields have more oscillations and do not smooth out over the metal/dielectric interface but present jumps. Besides, the electric field is larger than that for the 10 nm case in Fig. \ref{Efield}(a), since the effectively reduced nonlocal length scale implies that the plasmons have more time to travel towards the singularity while their wavelength is compressed, and confinement increases. Hence, higher field enhancements are to be found in metasurfaces of larger periods, which are surprisingly unaffected by radiative broadening.

\section{Conclusions}

In this paper we have explored the consequences of nonlocality for surfaces in the form of metasurfaces containing sharp edge singularities. Local theory predicts that external radiation will excite a continuous spectrum of modes with infinite energy density at the singularity. However, nonlocality forbids infinite concentration of charge at the surface with dramatic consequences for the spectra which are now discrete, and for the energy density which is no longer singular, although it takes a large value at the structural singularity. In effect nonlocality blunts sharp edges. This result suggests a way of measuring nonlocality, particularly by observing the mode spacing in the discrete spectrum it implies. When the metasurface period is very short compared to the free space wavelength of light, magnetism plays little part in the modes which are almost entirely electrostatic. In a local theory the spectrum is scale invariant hence the period of the metasurface does not affect the spectrum. This ceases to be true when nonlocality introduces a length scale in the form of the screening length. As a result, there is a balance between the nonlocal parameter, $\beta$, and the metasurface period: by increasing the period we can compensate for nonlocality at least until the period approaches the free space wavelength. Another point we noted was the insensitivity of radiative damping of the modes to the metasurface period. The charge currents generated by a mode couple weakly to external radiation and increasing the period might be expected to increase this coupling. For an isolated resonant particle this is indeed the case and radiative damping increases dramatically once the particle is much bigger than about 100 nm. In contrast, for a given length of metasurface, increasing the period means including less periods in the overall length hence the small effect.

\section{Methods}

\subsection{Field in the slab frame}
Since the conformal mapping conserves the spectrum of the system, we carry out our analytic calculation in the slab frame where the geometry is simpler. In the slab frame, we can write the field in each region by taking the quasi-static approximation as
{\scriptsize
\begin{equation}
H_z(k_x,y)=\left\{{\begin{array}{lr}
(1-r)a_a \frac{e^{-|k_x||y|}}{|k_x|} + (1+r)a_s \text{sgn}(k_x) \frac{e^{-|k_x||y|}}{\text{sgn}(y)|k_x|} + {b_ +}e^{-|k_x|y} + {b_ -}e^{|k_x|y},&{ -d_2 < y < d_1}\\
-t \frac{k_{0x}^{'}}{k_{0x}}  a_a \frac{e^{-|k_x||y + \frac{d}{2}|}}{|k_x|} -t a_s \text{sgn}(k_x) \frac{e^{-|k_x||y+\frac{d}{2}|}}{\text{sgn}(y+\frac{d}{2})|k_x|} + {c_ +}e^{-|k_x|y} + {c_ -}e^{|k_x|y},&{ -(d_2 + d_3) < y < - d_2}
\end{array}} \right.
\end{equation}}\normalsize
\begin{equation}
\varphi(k_x,y)=\left\{{\begin{array}{lr}
0,&{ -d_2 < y < d_1}\\
\frac{\mathrm{sgn}(k_x)}{\omega \varepsilon_0 \varepsilon}({d_ +}e^{-\kappa y} - {d_ -}e^{\kappa y}),&{ -(d_2 + d_3) < y < - d_2}
\end{array}} \right.
\end{equation}
where $H_z$ contributes to the transverse mode, and $\varphi$ to the longitudinal mode. Besides, the anti-symmetric source magnitude is $a_a = -i \frac{k_{0x}T}{4\pi}H_0$, while the symmetric one is $a_s = \frac{k_{0y}T}{4\pi}H_0$. The electric components can be obtained by 
\begin{equation}
\begin{split}
E_x(k_x,y) &= \frac{i}{\omega \varepsilon}\frac{\partial H_z}{\partial y} - \frac{\partial \varphi}{\partial x} \\
E_y(k_x,y) &= -\frac{i}{\omega \varepsilon}\frac{\partial H_z}{\partial x} - \frac{\partial \varphi}{\partial y}
\end{split} .
\end{equation}
Then by using the continuity of $H_z$, $E_x$ and $E_y$ at interface between metal and air, we can calculate the coefficients ($b_{\pm}$, $c_{\pm}$ and $d_{\pm}$) for the SPP mode in $k$-space. The anti-symmetric excitation gives us a mode with anti-symmetry, the dispersion relation for which is given in Eq. \ref{anti-symmetric mode dispersion}. In contrast, the symmetric excitation gives a symmetric mode, whose dispersion relation is given by Eq. \ref{symmetric mode dispersion}. For the anti-symmetric mode studied in this paper, the field distribution in real space ($H_z(x,y)$ and $\varphi(x,y)$) can be obtained by taking a Fourier transformation of the field in k-space($H_z(k_x,y)$ and $\varphi(k_x,y)$). The field distribution in real space can be written as
\begin{equation}
H_z^a(x,y)=\left\{{\begin{array}{lr}
\begin{split}
& i 2\pi a_a ({\Gamma_{a+}}e^{-\sqrt{k_{px}^2}y} + {\Gamma_{a-}}e^{\sqrt{k_{px}^2}y}) \\
& \times(e^{i\int_{0}^{|x|}k_{px}(x')d x'} + e^{-i\int_{0}^{|x|}k_{px}(x')d x'+i(\phi_0 + \phi)}) \frac{1}{1-e^{i(\phi_0 + \phi)}}
\end{split}
 ,&{ -d_2 < y < d_1}\\
\begin{split}
& i 2\pi a_a ({\Lambda_{a+}}e^{-\sqrt{k_{px}^2}y} + {\Lambda_{a-}}e^{\sqrt{k_{px}^2}y}) \\
& \times (e^{i\int_{0}^{|x|}k_{px}(x')d x'} + e^{-i\int_{0}^{|x|}k_{px}(x')d x'+i(\phi_0 + \phi)}) \frac{1}{1-e^{i(\phi_0 + \phi)}}
\end{split}
, &{ -(d_2 + d_3) < y < - d_2}
\end{array}} \right.
\end{equation}
\begin{equation}
\varphi^a(x,y)=\left\{{\begin{array}{lr}
0,&{ -d_2 < y < d_1}\\
\begin{split}
& \frac{i 2\pi a_a \mathrm{sgn}(k_{px}) \mathrm{sgn}(x)}{\omega \varepsilon_0 \varepsilon}({\Omega_{a+}}e^{-\kappa_p y} - {\Omega_{a-}}e^{\kappa_p y}) \\
& \times (e^{i\int_{0}^{|x|}k_{px}(x')d x'} - e^{-i\int_{0}^{|x|}k_{px}(x')d x'+i(\phi_0 + \phi)}) \frac{1}{1-e^{i(\phi_0 + \phi)}}
\end{split}
, &{ -(d_2 + d_3) < y < - d_2}
\end{array}} \right.
\end{equation}
where $\phi_0 = \int_{-L/2}^{L/2}k_{px}(x')d x'$ and $\phi$ is the phase change at the singularity which will be discussed in the next subsection. 

\subsection{Phase change at the singularity}
\begin{figure}[h]
\includegraphics[width=0.6\columnwidth]{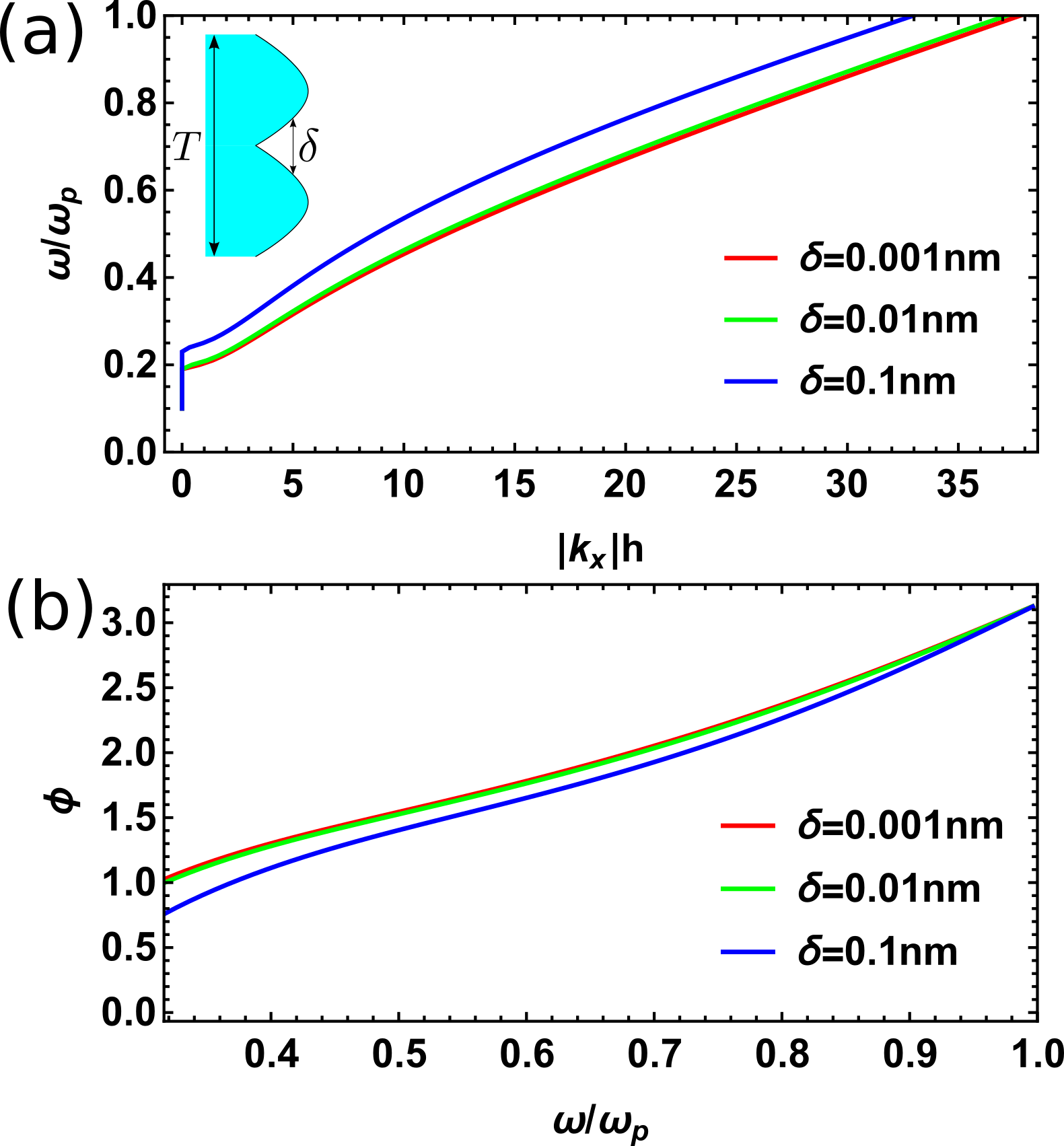}
\centering
\caption{$k$-vector and phase change $\phi$ at the singularity in the metasurface frame: (a)Dispersion relation ($h=T/2$ is the period of this cavity array) and (b) Phase change near the singularity for the cavity with different width $\delta$ (0.001 nm, 0.01 nm and 0.1 nm).}
\label{PhaseChange_groove}
\end{figure}

The WKB approximation fails near the turning point as $k$ is small\citep{griffiths2016introduction}. Therefore, we cannot use $k$ from the dispersion relation based on WKB to calculate the phase change $\phi$. Instead, the phase change at the tip is calculated in the metasurface frame. The result in Fig. \ref{PhaseChange_groove}(a) shows that the $k$-vector near the tip saturates as the width of the cavity $\delta$ decreases. The phase change $\phi$ of this cavity array can be obtained by matching the field at the terminus\citep{yang2018transformation, chandran2012metal, gordon2006light}. The calculated results show that both $k$-vector and phase change saturate near the singularity. Therefore, we use this value of the phase $\phi$ in the calculation.

\subsection{Reflection spectrum}
In order to calculate the reflection, we need to calculate the absorption in the slab frame. Our trick is to evaluate the power flow at the excitation point ($x=0$) by
\begin{equation}
P_{abs} = 2 \int_{-(d_2+d_3)}^{d_1} S_x dy
\end{equation}
where $S_x = \frac{1}{2} \mathrm{Re}[E_y^* H_z]$. The absorption by the slab is modeled as a real surface conductivity $\sigma_{er}$. Then using Kramers-Kronig relation \cite{yang2018transformation, jackson2012classical, dressel2005electrodynamics, kittel2004introduction}, a causal complex surface conductivity $\sigma_{e} = \sigma_{er} + i \sigma_{ei}$ gives the reflection spectrum of singular metasurface by
\begin{equation}
\begin{split}
 r = \frac{\sqrt{\epsilon}-1+\frac{\sigma_{e}}{\sigma_{e0}}}{\sqrt{\epsilon}+1+\frac{\sigma_{e}}{\sigma_{e0}}}
\end{split}.
\end{equation}

\subsection{Comsol modelling}
All of our numerical simulations are based on the RF and PDE modules in Comsol\citep{toscano2012modified, ciraci2012probing}, where the hydrodynamic system of equations in the metal are implemented as
\begin{equation}
\begin{split}
& \nabla \times \nabla \times \mathbf{E} = k_0^2 \mathbf{E} + i \omega \mu_0 \mathbf{J} \\
&\beta^2 \nabla (\nabla \cdot \mathbf{J}) + \omega(\omega+i \Gamma)\mathbf{J} = i\omega \omega_p^2 \varepsilon_0 \mathbf{E}
\end{split} 
\end{equation}
where the $\mathbf{J}$ is the current density for the electron. When $\beta \rightarrow 0$, we arrive at classical Drude local-response function $\varepsilon (\omega) = 1 - \omega_p^2/(\omega(\omega+i\Gamma))$. By solving the above coupled equations, the response of our singular metasurface can be obtained.
\begin{acknowledgement}
The authors thank A. I. Fernandez-Dominguez, Y. Luo and C. Ciraci for fruitful discussion. F.Y. acknowledges a Lee Family Scholarship for financial support. Y.T.W. acknowledges funding from the Leverhulme Trust. P.A.H and J.B.P. acknowledge funding from the Gordon and Betty Moore Foundation.
\end{acknowledgement}

%\begin{suppinfo}
%\end{suppinfo}


\bibliography{achemso-demo}

%\begin{figure}[h]
%\includegraphics{TOC.png}
%\centering
%\caption{For Table of Contents Only}
%\label{For Table of Contents Only}
%\end{figure}

\end{document}